\documentclass[a4paper,11pt]{article}
\usepackage{float}
\usepackage{jheppub} 
\usepackage{booktabs}
\usepackage{xcolor}
\usepackage{comment}
\usepackage{etoolbox}
\newcommand{\dd}{\mathop{}\!\mathrm{d}}

\usepackage{tikz}
\usetikzlibrary{arrows.meta,positioning}
\AtBeginEnvironment{thebibliography}{%
    \setlength{\itemsep}{3pt plus 0.2ex minus 0.05ex}%
}
\hypersetup{
    colorlinks=true,
    linkcolor=blue,
    citecolor=blue,
    urlcolor=blue
}

\title{\boldmath Holographic viscous transport in scalar-coupled Gauss–Bonnet gravity}
\author[a,b,c]{Tao Ma,}
\author[d,1]{Song He\note{Corresponding author.}}
\author[a,b,c,1]{and Li Li}

\affiliation[a]{School of Fundamental Physics and Mathematical
    Sciences, Hangzhou Institute for Advanced Study, UCAS, Hangzhou
    310024, China}
\affiliation[b]{Institute of Theoretical Physics, Chinese Academy of Sciences, Beijing 100190, China}
\affiliation[c]{University of Chinese Academy of Sciences, Beijing 100049, China}
\affiliation[d]{Institute of Fundamental Physics and Quantum Technology \& School of Physical Science and Technology, Ningbo University, Ningbo, Zhejiang 315211, China}

\emailAdd{matao24@mails.ucas.ac.cn, hesong@nbu.edu.cn, liliphy@itp.ac.cn}

\abstract{We test horizon formulas for the shear and bulk viscosities of a holographic plasma against viscosities extracted directly from retarded stress-tensor correlators. We work in a five-dimensional Einstein-dilaton-Gauss-Bonnet model calibrated to the lattice QCD equation of state at zero baryon density. UV-finite renormalized correlators are obtained by including the required curvature and scalar counterterms, and the fluctuation equations are solved while retaining the full dependence on the Gauss-Bonnet coupling $\alpha$, without using a horizon conservation law or an entropy-production formula. For constant coupling, the horizon conserved-current calculation reproduces the boundary Kubo results in both channels. For the scalar-dependent coupling studied here, both methods find a minimum in the shear-viscosity-to-entropy-density ratio $\eta/s$. In the bulk channel, the conserved-current calculation truncated at first order in \(\alpha \) yields a lower peak height in \(\zeta/s \), with the peak temperature nearly unchanged relative to the boundary result. For both couplings, the unmodified Eling-Oz (EO) formula, \(\zeta/s=(4\pi)^{-1}(\partial\phi_h/\partial\ln s)^2\), reproduces the boundary result for $\zeta/s$, where \(\phi_h\) is the horizon value of the scalar. The prefactor \(1/(4\pi)\) is thus retained even when the Gauss-Bonnet interaction shifts \(\eta/s\) from this value. These results provide a direct boundary benchmark for horizon prescriptions and support the unmodified EO formula beyond constant coupling.}


\begin{document}
\maketitle
\raggedbottom
\clubpenalty=10000
\widowpenalty=10000
\displaywidowpenalty=10000
\setlength{\textfloatsep}{14pt plus 2pt minus 2pt}
\setlength{\intextsep}{12pt plus 2pt minus 2pt}
\renewcommand{\topfraction}{0.90}
\renewcommand{\bottomfraction}{0.80}
\renewcommand{\textfraction}{0.10}
\renewcommand{\floatpagefraction}{0.75}

\section{Introduction}
\label{sec:introduction}
Hydrodynamics provides a universal effective description of interacting systems near local thermal equilibrium. It is an indispensable tool for understanding the quark-gluon plasma (QGP) created in heavy-ion collisions at RHIC and the
LHC~\cite{Bernhard:2019viscosities,JETSCAPE:2020mzn,Nijs:2021PRL,Nijs:2021Trajectum,Giacalone:2023NeutronSkin}. These experiments confirmed that the QGP behaves as a strongly coupled, nearly ideal fluid~\cite{Shuryak:2003xe,Shuryak:2004cy}. At first order in the gradient expansion, the plasma's dissipative response is specified by the shear and bulk viscosities, $\eta$ and $\zeta$. Both significantly affect hydrodynamic predictions and are increasingly constrained by Bayesian analyses of heavy-ion data~\cite{Bernhard:2019viscosities,JETSCAPE:2020mzn,Nijs:2021PRL,Nijs:2021Trajectum}.
However, their theoretical determination remains challenging. Perturbative QCD breaks down at strong coupling. Lattice QCD accurately constrains equilibrium thermodynamics~\cite{Borsanyi:2013bia,HotQCD:2014kol}. Viscosities, by contrast, require low-frequency real-time information that is difficult to extract from Euclidean correlators~\cite{Meyer:2011gj}.

Holography offers a powerful approach to strongly coupled transport. 
Two-derivative Einstein--dilaton models can reproduce nonconformal thermodynamics and the lattice equation of
state~\cite{Gubser:2008ny,Cai:2022omk,Fu:2024nmw,Rougemont:2023gfz,Jokela:2024xgz,Jarvinen:2025mgj,Shen:2025zkj,Zhu:2025gxo}. Such models also connect the dense-QCD phase structure with neutron-star
constraints~\cite{Ecker:2025vnb} and describe flavor-dependent thermodynamics~\cite{Chen:2024mmd,Zhao:2023gur,Jokela:2026jkh,He:2022amv,Liu:2023pbt}. In the isotropic setting, however, their shear-viscosity-to-entropy-density ratio is fixed at the universal, temperature-independent value $\eta/s=1/(4\pi)$~\cite{Policastro:2001yc,Kovtun:2004de}. Higher-curvature corrections can break this universality~\cite{Brigante:2007nu,Buchel:2009sk,Grozdanov:2016tdf}. In scalar
backgrounds, they can also generate temperature
dependence~\cite{Cremonini:2012ny,Buchel:2018aeu,Apostolidis:2025viscosities}.
A minimal realization couples a Gauss--Bonnet term to the scalar
field~\cite{Cai:2009zv,Tong:2026dwb}. Computing both viscosities in such models reliably is therefore important but challenging.

Horizon methods provide an efficient route to holographic transport by relating dissipative coefficients to quantities evaluated at the black-brane horizon~\cite{Kovtun:2003wp,Iqbal:2008by,Brustein:2008shear,Cai:2008effective,Paulos:2009transport}. Their application to higher-derivative gravity has, however, developed more extensively in the shear channel than in the bulk channel. Early studies considered conformal plasmas, for which the bulk viscosity vanishes~\cite{Myers:2009ij}, while calculations with a dilaton-dependent Gauss--Bonnet coupling focused on the shear viscosity and left the bulk channel unresolved~\cite{Cai:2009zv}. In Einstein--dilaton models, the bulk viscosity can be obtained from linearized fluctuations using the Gubser--Pufu--Rocha (GPR) method~\cite{Gubser:2008yx,Gubser:2008sz}. An alternative description is provided by the Eling--Oz (EO) formula, derived from horizon focusing and entropy production, whose agreement with the fluctuation calculation has been established in this setting~\cite{Eling:2011ms,Buchel:2011wx,Demircik:2023lsn}. At vanishing charge density, it takes the form
\begin{equation}
    \frac{\zeta}{s}
    =
    \frac{\eta}{s}
    \left(\frac{\partial\phi_h}{\partial\ln s}\right)^2 ,
    \label{eq:EO}
\end{equation}
where $\phi_h$ is the horizon value of the scalar field and $s$ is the entropy density; the derivative is taken at fixed scalar source. Since $\eta/s=1/(4\pi)$ in the two-derivative theory, this expression is equivalent to
\begin{equation}
    \frac{\zeta}{s}
    =
    \frac{1}{4\pi}
    \left(\frac{\partial\phi_h}{\partial\ln s}\right)^2 .
    \label{eq:EO-unmod}
\end{equation}
This equivalence is no longer automatic once higher-derivative interactions are included, since $\eta/s$ then deviates from $1/(4\pi)$. Consequently, the choice between Eqs.~\eqref{eq:EO} and~\eqref{eq:EO-unmod}, i.e., whether the prefactor is the corrected $\eta/s$ or the fixed value $1/(4\pi)$, is central to extending horizon formulas beyond Einstein gravity. A direct extension of the EO derivation to finite constant Gauss–Bonnet coupling yielded Eq.~\eqref{eq:EO-unmod} in the neutral, isotropic limit~\cite{Demircik:2024bxd}. In parallel, a radially conserved-current method, in the spirit of the GPR method~\cite{Gubser:2008yx,Gubser:2008sz}, was developed in Ref.~\cite{Buchel:2023fst} to compute both viscosities for curvature corrections with constant couplings. Generically, such corrections are higher-derivative, so their couplings must be treated perturbatively~\cite{Buchel:2024umq}. Within this perturbative treatment, the conserved-current calculation for the Gauss–Bonnet correction gave a bulk viscosity that differs from Eq.~\eqref{eq:EO}. Because the Gauss–Bonnet combination keeps the equations of motion second order, the same calculation was subsequently extended to finite coupling, and the resulting bulk viscosity was found to agree with Eq.~\eqref{eq:EO-unmod}~\cite{Buchel:2024umq}. In that work, Eq.~\eqref{eq:EO-unmod} was called the unmodified EO formula, and its validity was conjectured to be tied to the area law for the horizon entropy.

A scalar-dependent Gauss--Bonnet interaction raises further questions about this correspondence. Although scalar-dependent couplings have been incorporated into the conserved-current framework~\cite{Buchel:2026xtn}, the validity of the unmodified EO formula was not tested there. Independently, closed-form horizon expressions for both viscosities were derived from entropy production in scalar-dependent Gauss--Bonnet gravity~\cite{Tong:2025rxz}. The shear-viscosity expression agreed with an independent Kubo calculation, whereas the bulk-viscosity expression involved the negative bulk viscosity of the horizon membrane fluid, whose interpretation as a boundary transport coefficient requires care~\cite{Matsuo:2013qoa}. This membrane expression should be distinguished from the unmodified EO relation in Eq.~\eqref{eq:EO-unmod}. In a subsequent application of the conserved-current method of Ref.~\cite{Buchel:2023fst} to holographic QCD, the scalar-dependent bulk-channel calculation was performed only to linear order in the Gauss--Bonnet coupling $\alpha$~\cite{Tong:2026dwb}. The quantitative reliability of this truncation, as well as the applicability of the unmodified EO formula in the same model, therefore requires an independent test.

Boundary retarded correlators based on Kubo formulas can provide such a test, and they have been widely used in the context of low-energy many-body physics. In this work, a boundary Kubo framework is implemented in a five-dimensional Einstein--dilaton--Gauss--Bonnet (EDGB) model whose scalar potential is calibrated to lattice QCD thermodynamics~\cite{Tong:2026dwb}. The retarded stress-tensor correlators are extracted directly from the near-boundary response after holographic renormalization~\cite{deHaro:2000vlm,Skenderis:2002wp,Skenderis:2008dh}, and the shear and bulk viscosities are obtained from their Kubo formulas~\cite{Kubo:1957,Son:2002sd}. The full perturbation equations are solved without expanding in $\alpha$, and neither a horizon conservation law nor an entropy-production formula is used in the boundary extraction. To our knowledge, this provides the first fully renormalized boundary calculation of both viscosity channels in the scalar-dependent EDGB setting.

The main results are as follows:
\begin{itemize}
    \item Finite retarded correlators are obtained in both the shear and bulk channels. The counterterms used for equilibrium thermodynamics in Refs.~\cite{Cai:2022omk,Tong:2026dwb,Tong:2025rxz} are supplemented by the curvature and scalar counterterms required to renormalize the stress-tensor two-point functions. Both $\eta/s$ and $\zeta/s$ are then extracted while retaining the full dependence on $\alpha$.

    \item The boundary results agree with the conserved-current results of Ref.~\cite{Tong:2026dwb} in the shear channel and in the constant-coupling bulk channel. For scalar-dependent coupling, the finite-$\alpha$ boundary result and the $\mathcal{O}(\alpha)$ conserved-current result exhibit similar temperature dependence and nearly identical peak positions, but different peak heights. This difference is interpreted as an effect of the linear-order truncation in the latter calculation.

    \item The boundary bulk viscosity agrees numerically with the unmodified EO formula~\eqref{eq:EO-unmod} for both constant coupling and the scalar-dependent coupling studied here. This provides direct boundary evidence for the prefactor $1/(4\pi)$ in place of the corrected ratio $\eta/s$. Since the planar horizon entropy obeys the area law in both cases, the agreement also supports the conjecture of Ref.~\cite{Buchel:2024umq} beyond constant coupling.

    \item The resulting viscosity ratios are compared with the JETSCAPE Bayesian credible intervals~\cite{JETSCAPE:2020mzn}. The temperature-dependent agreement and deviations indicate the phenomenological scope of the benchmark and motivate further refinement of the coupling function and model parameters.
\end{itemize}

The boundary formulation also provides a basis for applications beyond the present holographic QCD model. By organizing the calculation in terms of physical sources and renormalized responses, it offers a systematic route to retarded correlators and the associated real-time observables. The framework can be extended to multiple fields, charged systems, and other higher-derivative theories. The present calculation provides a concrete implementation in which we can test scalar--metric mixing, boundary renormalization, and finite Gauss--Bonnet coupling together.

The remainder of this paper is organized as follows. Section~\ref{sec:holographic_setup} introduces the model and the boundary Kubo framework. Section~\ref{sec:viscosity_calculation} outlines the counterterms and the boundary extraction of both viscosities. Section~\ref{sec:horizon_validation} compares our results with those from the conserved-current method and the unmodified EO formula~\eqref{eq:EO-unmod}. Section~\ref{sec:conclusion} concludes with a brief outlook. We collect technical details on the background equations and asymptotics, holographic renormalization, the derivation of both viscosities, and numerical calculations in Appendices~\ref{app:background}, \ref{sec:B1}, \ref{sec:C}, and~\ref{app:numerical_details}, respectively.

\section{Holographic setup}
\label{sec:holographic_setup}

\subsection{Action and background geometry}
\label{sec:action_background}
We begin with the EDGB model, defined by the action
\begin{align}
S_{\rm bulk}=\dfrac{1}{2\kappa_5^2}\int \dd^5 x\sqrt{-g}\left[
\mathcal{R}-\dfrac{1}{2}\nabla_\mu \phi\nabla^\mu \phi
-V(\phi)+\alpha H(\phi)\mathcal{G}
\right],
\label{eq2.1}
\end{align}
where $\kappa_5$ denotes the five-dimensional gravitational coupling constant,
$\mathcal{R}$ is the Ricci scalar associated with the metric $g_{\mu\nu}$,
and $\alpha$ controls the strength of the higher-curvature correction. 
The Gauss--Bonnet term is defined as~\cite{Lovelock:1971yv}
\begin{align}
    \mathcal{G}=\mathcal{R}^2-4 \mathcal{R}_{\mu \nu} \mathcal{R}^{\mu \nu}
    +\mathcal{R}_{\mu \nu \lambda \rho} \mathcal{R}^{\mu \nu \lambda \rho}.
\end{align}
The coupling function $H(\phi)$ controls the scalar dependence of the Gauss--Bonnet coupling. We adopt the form used in Ref.~\cite{Tong:2026dwb},
\begin{equation}
    H(\phi)=1-\frac{h_1 \phi^6}{h_2+\phi^6},
    \label{eq:H_nonconstant}
\end{equation}
so that $H(\phi)=1+\mathcal{O}(\phi^6)$ as $\phi\to 0$. The scalar potential in the action~\eqref{eq2.1} is taken from the same work,
\begin{equation}
    V(\phi)=-12\cosh(c_1\phi)
    +\left(6c_1^2-\frac{3}{2}\right)\phi^2
    +c_2\phi^3+c_3\phi^6+24\alpha.
    \label{eq:potential}
\end{equation}
The parameters $c_1$, $c_2$, $c_3$ and $h_1$, $h_2$ are fixed by its lattice equation-of-state calibration.

For a static, isotropic thermal state on Minkowski spacetime, we consider the planar black-brane metric
\begin{equation}\label{eq2.2}
     \dd s^2=-f(r) e^{-\eta(r)} \dd t^2+\frac{\dd r^2}{f(r)}+r^2\left(\dd x^2+\dd y^2+\dd z^2\right),
\end{equation}
with a purely radial scalar field $\phi=\phi(r)$. Varying the action~\eqref{eq2.1} and substituting this ansatz yields the equations of motion, whose explicit forms are listed in Appendix~\ref{sec:A.1}.

We require the scalar to vanish as $r\to\infty$, so that the geometry approaches AdS. This is consistent because $\phi=0$ is an extremum of the effective potential $V(\phi)-\alpha H(\phi)\mathcal{G}$, since $V'(0)=H'(0)=0$ for Eqs.~\eqref{eq:H_nonconstant} and~\eqref{eq:potential}.
The constant $24\alpha$ in Eq.~\eqref{eq:potential} sets $V(0)=-12+24\alpha$ and keeps the effective AdS radius at unity. Since $H(\phi)-1=\mathcal{O}(\phi^6)$, the Gauss--Bonnet term does not shift the effective scalar mass, $m_{\rm eff}^2=-3$. This gives the scaling dimension $\Delta=3$ of the dual operator via $m_{\rm eff}^2=\Delta(\Delta-4)$. The two independent modes of the scalar then fall off as
\begin{equation}
    \phi(r)=\frac{\phi_s}{r}+\cdots+\frac{\phi_v}{r^3}+\cdots,
\end{equation}
where the source $\phi_s$ of the dual operator explicitly breaks conformal symmetry, and $\phi_v$ encodes its response. The full UV expansions are given in Appendix~\ref{app:IR_UV_expansions}.

We denote the event horizon by $r_h$, where the blackening function vanishes, $f(r_h)=0$. The Hawking temperature is given by
\begin{equation}
    T=\frac{1}{4\pi}f^{\prime}(r_h)e^{-\eta(r_h)/2}.
\label{eq:Hawking_temperature}
\end{equation}
The entropy density $s$ is given by the Wald formula~\cite{Iyer:1994ys}. For this planar black brane, the Gauss--Bonnet contribution to the Wald entropy is identically zero, leading to~\cite{Tong:2025rxz}
\begin{equation}
    s=\frac{2\pi r_h^3}{\kappa_5^2}.
\label{eq:entropy_density}
\end{equation}

\subsection{Boundary Kubo framework}
\label{sec:uv_framework}
We first introduce the general boundary Kubo framework~\cite{Son:2002sd,Kaminski:2009dh}. For bulk fluctuations dual to boundary operators $\mathcal O_A$, we eliminate gauge redundancies and solve the constraint equations. We then impose infalling boundary conditions at the horizon and integrate the perturbation equations to the boundary. The coefficients of the independent near-boundary modes encode the boundary sources and the corresponding responses. Holographic renormalization removes UV divergences of the on-shell action via boundary counterterms~\cite{deHaro:2000vlm,Skenderis:2002wp}, yielding finite responses. The linear response to the sources then gives the retarded two-point functions. Finally, we extract physical observables, such as transport coefficients, from the Kubo formulas~\cite{Kubo:1957}. Figure~\ref{fig:boundary_response_flowchart} summarizes this workflow:
\begin{figure}[H]
    \centering
    \begin{tikzpicture}[
      node distance=11mm and 22mm,
      stage/.style={
        draw=black!58,
        fill=black!2,
        rounded corners=1.2pt,
        line width=.45pt,
        text width=30mm,
        minimum height=17mm,
        align=center,
        inner xsep=2pt,
        inner ysep=2.6pt,
        font=\small
      },
      flow/.style={
        -{Triangle[length=2.2mm,width=1.8mm]},
        draw=black,
        line width=.9pt
      }
    ]
    \node[stage] (bulk) {%
      \textbf{1. Bulk}\\[-1pt]
      \textbf{fluctuations}\\[2pt]
      $\delta\Psi_A$};
    \node[stage,right=22mm of bulk] (uv) {%
      \textbf{2. UV data}\\[2pt]
      $\{J_A,q_A\}$};
    \node[stage,right=22mm of uv] (action) {%
      \textbf{3. On-shell}\\[-1pt]
      \textbf{action}\\[2pt]
      $S_{\mathrm{ren}}^{\mathrm{on\text{-}shell}}$};
    \node[stage,below=11mm of action] (response) {%
      \textbf{4. Response}\\[2pt]
      $\langle\mathcal O_A\rangle_{\mathrm{ren}}$};
    \node[stage,left=22mm of response] (green) {%
      \textbf{5. Retarded}\\[-1pt]
      \textbf{Green function}\\[2pt]
      $G^R_{AB}$};
    \node[stage,left=22mm of green] (observable) {%
      \textbf{6. Physical}\\[-1pt]
      \textbf{observables}\\[2pt]
      $\eta,\,\zeta,\,\dots$};
    \draw[flow] (bulk) -- node[midway,above=.6mm,font=\scriptsize,align=center]{Infalling\\[-1pt]condition} (uv);
    \draw[flow] (uv) -- node[midway,above=.6mm,font=\scriptsize,align=center]{Holographic\\[-1pt]renormalization} (action);
    \draw[flow] (action) -- node[midway,right=.8mm,font=\scriptsize,align=left]{Functional\\[-1pt]derivative} (response);
    \draw[flow] (response) -- node[midway,above=.6mm,font=\scriptsize,align=center]{Linear response} (green);
    \draw[flow] (green) -- node[midway,above=.6mm,font=\scriptsize,align=center]{Kubo formulas} (observable);
    \end{tikzpicture}
    \caption{Schematic of the boundary Kubo framework. Bulk fluctuations are solved with infalling boundary conditions and matched to UV data; holographic renormalization yields the on-shell action and the renormalized response. The retarded Green's function follows from linear response, and physical observables such as $\eta$ and $\zeta$ are obtained via the Kubo formulas.}
    \label{fig:boundary_response_flowchart}
\end{figure}
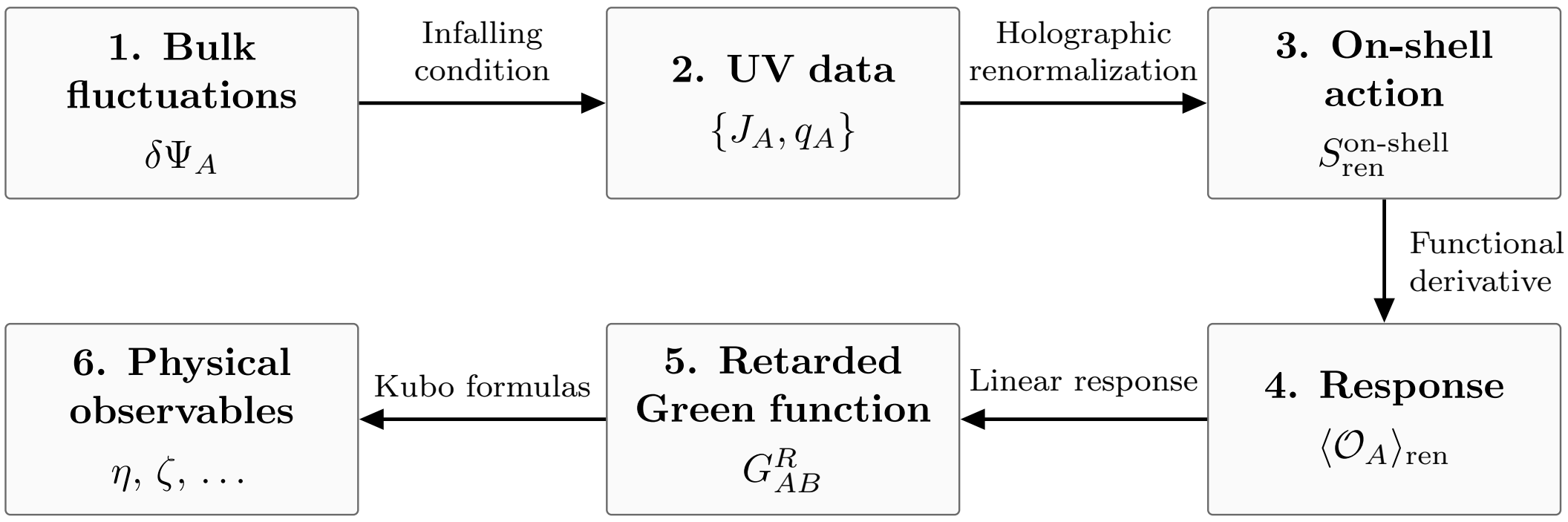

To make this framework concrete, we demonstrate its implementation within the EDGB model introduced above. Working in Fourier space with the convention $e^{-i\omega t+i\mathbf{k}\cdot\mathbf{x}}$, we decompose the general solution for a single decoupled channel near the boundary as
\begin{align}
\Psi_A(r;\omega,\mathbf{k})
\sim
J_A(\omega,\mathbf{k})\Psi_A^{(s)}(r)
+
q_A(\omega,\mathbf{k})\Psi_A^{(v)}(r),
\qquad r\to\infty.
\label{eq:general_uv_source_response}
\end{align}
Here, $\Psi_A^{(s)}$ and $\Psi_A^{(v)}$ denote the source and response branches, respectively, each including its full near-boundary expansion. $J_A$ serves as the external source for the dual operator $\mathcal O_A$, while $q_A$ encodes its response. Note that Eq.~\eqref{eq:general_uv_source_response} specifies only the formal near-boundary asymptotic structure; extracting the physical renormalized response $\left\langle\mathcal O_A\right\rangle_{\mathrm{ren}}$ requires holographic renormalization of the on-shell action. Thus, the renormalized one-point function
$\langle\mathcal O_A\rangle_{\mathrm{ren}}$ is defined by
\begin{align}
\left\langle\mathcal O_A\right\rangle_{\mathrm{ren}}
=
\frac{\delta S_{\mathrm{ren}}^{\mathrm{on\text{-}shell}}}
{\delta J_A},
\label{eq:general_ren_variation}
\end{align}
where $S_{\mathrm{ren}}^{\mathrm{on\text{-}shell}}$ is the renormalized on-shell action. For a boundary metric source, the response is the renormalized stress tensor,
whose normalization is fixed by varying the renormalized on-shell action with
respect to the boundary metric (see Appendix~\ref{sec:B1}).

Following Refs.~\cite{Son:2002sd, Kovtun:2012rj}, the retarded correlator in Minkowski space is defined as
\begin{align}
G^R_{AB}(\omega,\mathbf{k})
=
-i\int \dd^4x\,e^{i\omega t-i\mathbf{k}\cdot\mathbf{x}}\theta(t)
\left\langle
[\mathcal O_A(t,\mathbf{x}),\mathcal O_B(0,\mathbf{0})]
\right\rangle.
\label{eq:retarded_convention}
\end{align}
For the action coupling convention $+J_B\mathcal O_B$, the corresponding
renormalized correlator is obtained from the linear response relation
\begin{align}
G^R_{AB}(\omega,\mathbf{k})
=
-\left.
\frac{\delta
\left\langle\mathcal O_A(\omega,\mathbf{k})\right\rangle_{\mathrm{ren}}}
{\delta J_B(\omega,\mathbf{k})}
\right|_{J=0}.
\label{eq:general_single_response}
\end{align}
Here, $J=0$ denotes evaluation at vanishing perturbative sources, with the equilibrium background held fixed. For a single decoupled channel, this expression reduces to the renormalized response-to-source ratio, with the overall minus sign fixed by the convention above.

For coupled perturbations~\cite{Kaminski:2009dh}, we assemble the near-boundary
source and response coefficients of independent infalling solutions into the
columns of $\mathbf S$ and $\mathbf R$, respectively. When $\mathbf S$ is
invertible, the renormalized retarded correlator takes the form
\begin{align}
\mathbf G^R(\omega,\mathbf{k})
=
\mathbf N\,\mathbf R\mathbf S^{-1}+\mathbf C.
\label{eq:general_matrix_response}
\end{align}
Here, $\mathbf N$ accounts for the normalization and sign conventions, and
$\mathbf C$ collects local contact terms. For the bulk channel considered below,
however, a gauge-invariant master
field~\cite{Gubser:2008sz, Buchel:2023fst, Tong:2025rxz} reduces the calculation
to a single response-to-source ratio for a spatial-trace metric source.

\section{Boundary extraction of viscosities}
\label{sec:viscosity_calculation}

We now evaluate the shear viscosity $\eta$ and the bulk viscosity $\zeta$ with the
boundary Kubo framework. We first establish the boundary counterterm action that
renormalizes the two-point functions, and then derive the fluctuation equations
and Kubo formulas for both channels. Complete derivations are given in
Appendices~\ref{sec:B1} and~\ref{sec:C}.

\subsection{Holographic counterterms}
\label{Counterterms}
{To obtain a finite stress tensor, the bulk action~\eqref{eq2.1} must be supplemented with boundary counterterms. The counterterms of Refs.~\cite{Tong:2025rxz,Tong:2026dwb,Cai:2022omk} render the background stress tensor finite. However, the linear response of the stress tensor, which gives the two-point functions, requires additional intrinsic-curvature and scalar counterterms. These vanish on the background but contribute at linear order in the perturbations. The full counterterm action reads
\begin{align}
S_{\rm ct}
={}&
S_{\rm ct}^{(0)}
-\frac{1}{2\kappa_5^2}
\int_{\partial\mathcal{M}} \dd^4x\sqrt{-\gamma}\,
\bigg[
\left(\frac{1}{2}+2\alpha\right)\widehat R
+\left(\frac{1}{4}-\alpha\right)
\left(\widehat R_{\mu\nu}\widehat R^{\mu\nu}-\frac{1}{3}\widehat R^2\right)\ln r
\nonumber\\
&\hspace{3.2cm}
+\frac{1}{2}\left(\phi\widehat\Box\phi-\frac{1}{6}\widehat R\phi^2\right)\ln r
\bigg],
\label{eq:Sct_summary}
\end{align}
where $S_{\rm ct}^{(0)}$ is the counterterm action of Refs.~\cite{Tong:2025rxz,Tong:2026dwb,Cai:2022omk}, given in Eq.~\eqref{eq:Sct_flat_def}. Here $\gamma_{\mu\nu}$ is the induced metric on the cutoff surface $\partial\mathcal{M}$, with $\gamma=\det\gamma_{\mu\nu}$. All hatted quantities are built from $\gamma_{\mu\nu}$, and $\widehat\Box\equiv\gamma^{\mu\nu}\widehat\nabla_\mu\widehat\nabla_\nu$. The full boundary action and the renormalized stress tensor are given in Eqs.~\eqref{eq:Sboundary_general_H} and~\eqref{eq:Tmunu_general_H}.}

\subsection{Shear channel}
\label{sec:shear_viscosity}

In the shear channel, we consider an off-diagonal metric perturbation at zero spatial momentum, $\mathbf{k}=\mathbf{0}$,
\begin{align}
    \delta g_{xy}(t,r)=r^2 h(r)e^{-i\omega t},
\label{eq:shear_perturbation}
\end{align}
where $h(r)$ is the radial profile and $\omega$ is the frequency. The linearized equation for $h(r)$ takes the compact form
\begin{align}
h''(r)
+\left[
\frac{3}{r}+\frac{f'}{f}-\frac{\eta'}{2}
+\mathcal P_{\mathrm{GB}}(r)
\right]h'(r)
+\omega^2\left[
\frac{e^\eta}{f^2}
+\mathcal Q_{\mathrm{GB}}(r)
\right]h(r)=0 ,
\label{eq:shear_master}
\end{align}
where $\mathcal P_{\mathrm{GB}}(r)$ and $\mathcal Q_{\mathrm{GB}}(r)$ encode the Gauss--Bonnet corrections, depending on $f(r)$, $\eta(r)$, $\phi(r)$ and the coupling $H(\phi)$. Their explicit forms are given in Eqs.~\eqref{eq:shear_P_GB} and~\eqref{eq:shear_Q_GB}. Near the UV boundary, $h(r)$ has the asymptotic expansion
\begin{align}
h(r)
={}&
h_s
+\frac{h_s\omega^2}{4r^2}
+\frac{h_v}{r^4}
-\frac{h_s\phi_s^2\omega^2}{48(1-4\alpha)r^4}\ln r
+\frac{h_s\omega^4}{16r^4}\ln r
\nonumber\\
&
-\frac{h_s c_2\phi_s^3\omega^2}{15(1-4\alpha)r^5}
+\mathcal{O}\!\left(\frac{\ln r}{r^6}\right),
\qquad r\to\infty .
\label{eq:shear_uv_expansion}
\end{align}
Here, $h_s$ and $h_v$ are the source and response coefficients of the metric perturbation, respectively.

{The shear viscosity is determined by the Kubo formula~\cite{Kubo:1957,Son:2002sd},
\begin{equation}
\eta=-\lim_{\omega\to0}\frac{1}{\omega}
\mathrm{Im}\,G_{\eta}^R(\omega,\mathbf{k}=\mathbf{0}),
\label{eq:kubo_eta}
\end{equation}
where $G^R_\eta$ is the retarded correlator~\eqref{eq:retarded_convention} of $T_{xy}$, the $xy$ component of the renormalized stress tensor~\eqref{eq:Tmunu_general_H}. We impose the infalling condition $h(r)\sim(r-r_h)^{-i\omega/(4\pi T)}$ near the horizon and read off $G^R_\eta$ from the linear response of $T_{xy}$.
Dividing Eq.~\eqref{eq:kubo_eta} by the entropy density $s$ then gives}
\begin{align}
\frac{\eta}{s}
=\frac{1-4\alpha}{\pi r_h^3}
\lim_{\omega\to0}\frac{1}{\omega}
\mathrm{Im}\left(\frac{h_v}{h_s}\right).
\label{eq:eta_over_s_uv}
\end{align}

\subsection{Bulk channel}
\label{sec:bulk_viscosity}
In the bulk channel, we consider $\mathrm{SO}(3)$-invariant metric and scalar perturbations at zero spatial momentum, $\mathbf{k}=\mathbf{0}$~\cite{Buchel:2023fst,Tong:2025rxz}. Diffeomorphism invariance permits the choice of the radial gauge $\delta g_{r\mu}=0$ ($\mu=t,r,x,y,z$)~\cite{Gubser:2008sz,Buchel:2023fst}. The remaining metric perturbations are 
\begin{align}
\begin{aligned}
\delta g_{tt} &= -f(r) e^{-\eta(r)} h_{00}(r) e^{-i\omega t}, \\
\delta g_{xx} &= \delta g_{yy} = \delta g_{zz} = r^2 h_{11}(r) e^{-i\omega t},
\end{aligned}
\end{align}
where $h_{00}(r)$ and $h_{11}(r)$ denote the temporal and spatial-trace metric modes, respectively, up to residual gauge redundancy.

Extracting bulk viscosity requires a spatial-trace metric source at the boundary. In the spacetime interior, however, the metric perturbations inevitably excite scalar fluctuations $\delta\phi(r,t)=\delta\phi(r)e^{-i \omega t}$. Consequently, the metric and scalar modes become coupled. Moreover, both modes carry residual gauge redundancy. To remove this gauge freedom and decouple the system, we adopt the master field~\cite{Buchel:2023fst,Tong:2025rxz}
\begin{align}
Z(r) = \delta\phi(r) - \frac{r\phi'(r)}{2}h_{11}(r).
\label{eq:Zphi_definition}
\end{align}
By construction, $Z(r)$ is gauge invariant and isolates the single propagating degree of freedom, as demonstrated in Appendix~\ref{sec:C_bulk}. Therefore, we only need to compute a single response-to-source ratio rather than the coupled matrix.

Substituting Eq.~\eqref{eq:Zphi_definition} into the linearized equations to eliminate $h_{00}(r)$ and $h_{11}(r)$, the master equation for $Z(r)$ takes the form
\begin{align}
Z''(r)
&+\left[
\frac{3}{r}+\frac{f'}{f}-\frac{\eta'}{2}
+\frac{K_{Z,r}'}{K_{Z,r}}
\right]Z'(r)
+\left[
\omega^2\frac{e^\eta}{f^2}\frac{K_{Z,t}}{K_{Z,r}}
-\frac{\mathcal V_Z}{fK_{Z,r}}
\right]Z(r)=0 .
\label{eq:bulk_master}
\end{align}
Here, $K_{Z,r}(r)$ and $K_{Z,t}(r)$ are the radial and temporal factors, while $\mathcal V_Z(r)$ is the effective potential. Their complete expressions, including all corrections generated by $\alpha H(\phi)\mathcal G$, are given in Eqs.~\eqref{eq:bulk_common_factors}--\eqref{eq:bulk_effective_potential} of Appendix~\ref{sec:C_bulk}. 

Near the asymptotically AdS boundary, $Z(r)$ has the expansion
\begin{align}
Z(r)
={}&
\frac{Z_s}{r}
-\frac{6c_2\phi_s Z_s}{r^2}
+\frac{Z_v}{r^3}
-\frac{Z_s\phi_s^2}{2(1-4\alpha)r^3}
\left[1-(1-4\alpha)\left(54c_2^2+6c_1^4\right)\right]\ln r
\nonumber\\
&+\frac{Z_s\omega^2}{2r^3}\ln r
+\mathcal{O}\!\left(\frac{\ln r}{r^4}\right),
\qquad r\to\infty.
\label{eq:Zphi_uv_bulk}
\end{align}
Here, the leading coefficient $Z_s$ represents the boundary source, while the subleading coefficient $Z_v$ encodes the response.

For the bulk viscosity, the Kubo formula reads~\cite{Martinez:2019bsn}
\begin{align}
\zeta=-\frac{1}{9}\lim_{\omega\to0}\frac{1}{\omega}\mathrm{Im}\,G^R_{\zeta}(\omega,\mathbf{k}=\mathbf{0}),
\label{eq:kubo_zeta}
\end{align}
where $G^R_{\zeta}$ is the retarded correlator~\eqref{eq:retarded_convention} of the spatial trace $\mathcal{T}=T_{xx}+T_{yy}+T_{zz}$. Here $T_{xx}$, $T_{yy}$, and $T_{zz}$ are the diagonal spatial components of the renormalized stress tensor~\eqref{eq:Tmunu_general_H}.
We likewise impose the infalling condition $Z(r)\sim(r-r_h)^{-i\omega/(4\pi T)}$ at the horizon. The linear response of $\mathcal T$ determines $G^R_\zeta$. Normalized by the entropy density $s$, Eq.~\eqref{eq:kubo_zeta} becomes
\begin{align}
\frac{\zeta}{s}
=\frac{\phi_s^2}{18\pi r_h^3}
\lim_{\omega\to0}\frac{1}{\omega}
\mathrm{Im}\!\left(\frac{Z_v}{Z_s}\right).
\label{eq:zeta_over_s_uv}
\end{align}
See Appendix~\ref{sec:C_bulk} for further details.

\section{Validation against horizon prescriptions}
\label{sec:horizon_validation}
We compare the boundary Kubo results in Eqs.~\eqref{eq:eta_over_s_uv} and~\eqref{eq:zeta_over_s_uv} with those of existing horizon prescriptions for the
constant and scalar-dependent Gauss--Bonnet couplings. Specifically, we benchmark both channels against Ref.~\cite{Tong:2026dwb}, which applies the conserved-current method of Ref.~\cite{Buchel:2023fst}. For bulk viscosity, we also compare against the unmodified EO formula~\eqref{eq:EO-unmod}, evaluated numerically as in~\eqref {eq:EO_numerical_working} in Appendix~\ref{app:eo_numerical}.

\subsection{Constant coupling: $H(\phi)=1$}
\label{sec:horizon_constant}
Setting $h_1=0$ in Eq.~\eqref{eq:H_nonconstant}, namely $H(\phi)=1$, yields the constant-coupling case. The model parameters from Ref.~\cite{Tong:2026dwb} are listed in Table~\ref{tab:model_parameters}. Here, $b$ denotes the holographic renormalization parameter (see Appendix~\ref{sec:B1}).

\begin{table}[H]
  \centering
  \caption{Model parameters for the constant coupling $H(\phi)=1$}
  \label{tab:model_parameters}
  \renewcommand{\arraystretch}{1.2}
  \setlength{\tabcolsep}{12pt}
  \begin{tabular}{@{}ccccccc@{}}
    \toprule
    $\alpha$ & $c_1$ & $c_2$ & $c_3$
      & $\kappa_5^2$ & $b$ & $\phi_s $ \\
    \midrule
    $-0.006$ & $0.7074$ & $0$ & $0.0041$
    & $2\pi(1.62)$ & $-0.2744$ & $1065$ \\
    \bottomrule
  \end{tabular}
\end{table}

\begin{figure}[!tbp]
\centering
    \includegraphics[width=0.76\textwidth]{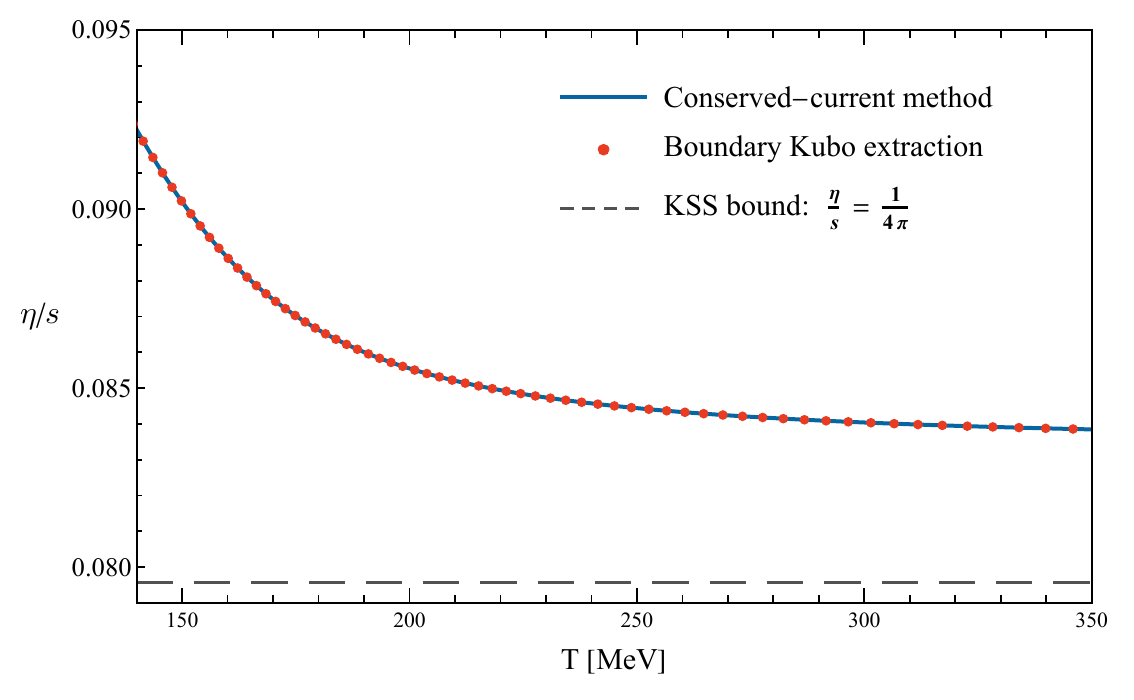}
    \caption{Shear-viscosity-to-entropy-density ratio $\eta/s$ as a function of temperature for the constant Gauss--Bonnet coupling $H(\phi)=1$ with $\alpha=-0.006$. Red circles: boundary Kubo extraction; blue solid curve: conserved-current result of Ref.~\cite{Tong:2026dwb}. The two agree to graphical accuracy over the plotted temperature range, and both lie above the KSS bound $1/(4\pi)$ (black dashed line)~\cite{Kovtun:2003wp,Kovtun:2004de}.}
    \label{fig:eta_constant_uv_horizon}

\vspace{14pt}

\centering
    \includegraphics[width=0.76\textwidth]{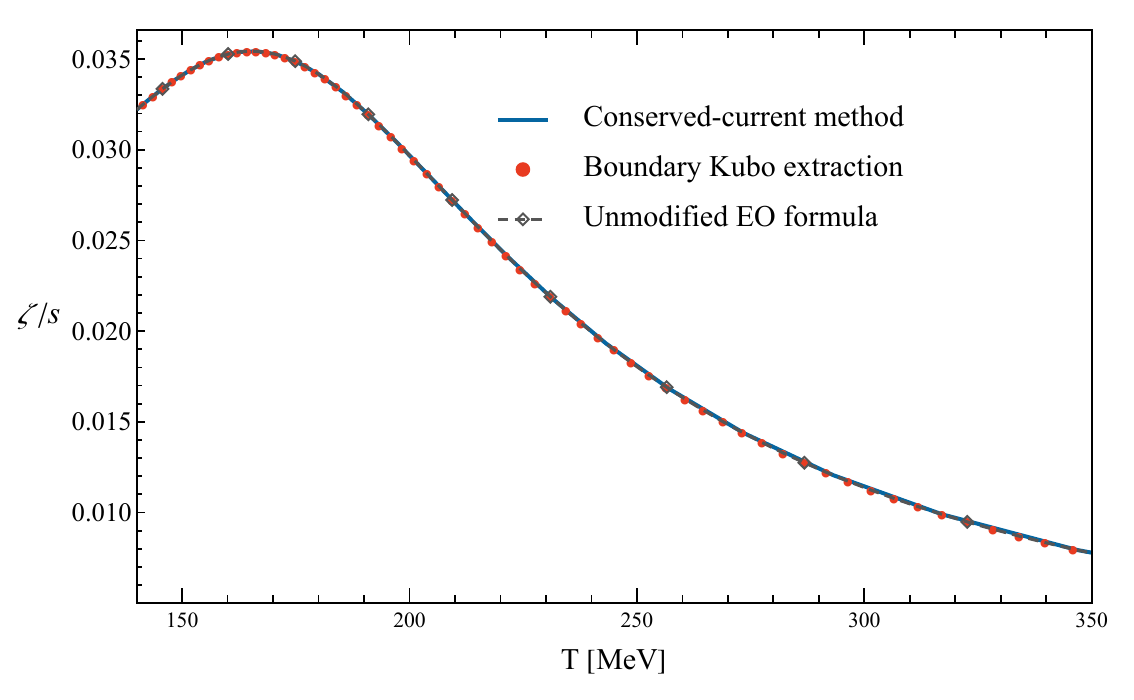}
    \caption{Bulk-viscosity-to-entropy-density ratio $\zeta/s$ versus temperature for the constant Gauss--Bonnet coupling $H(\phi)=1$ with $\alpha=-0.006$. Red circles: boundary Kubo result; blue solid curve: conserved-current result of Ref.~\cite{Tong:2026dwb}; gray dashed curve with open diamonds: unmodified EO formula, Eq.~\eqref {eq:EO-unmod}. All three agree to graphical accuracy over the plotted temperature range.}
    \label{fig:zeta_constant_uv_horizon}
\end{figure}

Figure~\ref{fig:eta_constant_uv_horizon} compares the conserved-current result of
Ref.~\cite{Tong:2026dwb} with our result from Eq.~\eqref{eq:eta_over_s_uv} in the
shear channel. At graphical resolution, the two agree throughout the plotted
temperature range. Both give a smooth, monotonically decreasing $\eta/s$ that
stays above the KSS bound $1/(4\pi)$~\cite{Kovtun:2003wp,Kovtun:2004de}.

For bulk viscosity with constant coupling, Figure~\ref{fig:zeta_constant_uv_horizon} compares three independent approaches: the boundary Kubo framework, the conserved-current method of Ref.~\cite{Tong:2026dwb}, and the unmodified EO formula~\eqref{eq:EO-unmod}. All three curves agree over the entire temperature range. This agreement has three
implications:
\begin{itemize}
    \item It provides a nontrivial cross-check of our framework. The agreement tests the full calculation chain, from the finite-$\alpha$ equation for $Z(r)$ and its infalling solution to the extraction of boundary response.

    \item It independently verifies the unmodified EO formula in constant-coupling Gauss--Bonnet gravity. The prefactor remains $(4\pi)^{-1}$ rather than the Gauss--Bonnet-corrected $\eta/s$, i.e., Eq.~\eqref{eq:EO-unmod} holds rather than Eq.~\eqref{eq:EO}. This is consistent with the direct EO derivation of Ref.~\cite{Demircik:2024bxd} and the conserved-current result of Ref.~\cite{Buchel:2024umq}.
    \item All three approaches capture the same nonconformal peak in $\zeta/s$ near the crossover. This agreement establishes a common baseline for the scalar-dependent case below.
\end{itemize}

\subsection{Scalar-dependent coupling: \texorpdfstring{$H(\phi)\neq 1$}{H(phi) != 1}}
\label{sec:horizon_nonconstant}
For the scalar-dependent coupling, we choose the benchmark with $h_1=h_2=0.5$ in Eq.~\eqref{eq:H_nonconstant},
\begin{align}
 h_1=h_2=0.5,\quad
 H(\phi)=1-\frac{\phi^6}{1+2\phi^6}.
 \label{eq:H_nonconstant_benchmark}
\end{align}
The corresponding model parameters, adopted from Ref.~\cite{Tong:2026dwb}, are summarized in Table~\ref{tab:model_parameters_nonconstant}:

\begin{table}[H]
  \centering
  \caption{Model parameters for the scalar-dependent coupling $H(\phi)=1-\phi^6/(1+2\phi^6)$}
  \label{tab:model_parameters_nonconstant}
  \renewcommand{\arraystretch}{1.2}
  \setlength{\tabcolsep}{12pt}
  \begin{tabular}{@{}ccccccc@{}}
    \toprule
    $\alpha$ & $c_1$ & $c_2$ & $c_3$
      & $\kappa_5^2$ & $b$ & $\phi_s$ \\
    \midrule
    $-0.008$ & $0.733$ & $0.10957$ & $0.0046$
    & $2\pi(1.49)$ & $-0.2744$ & $985$  \\
    \bottomrule
  \end{tabular}
\end{table}

\begin{figure}[!tbp]
\centering
    \includegraphics[width=0.68\textwidth]{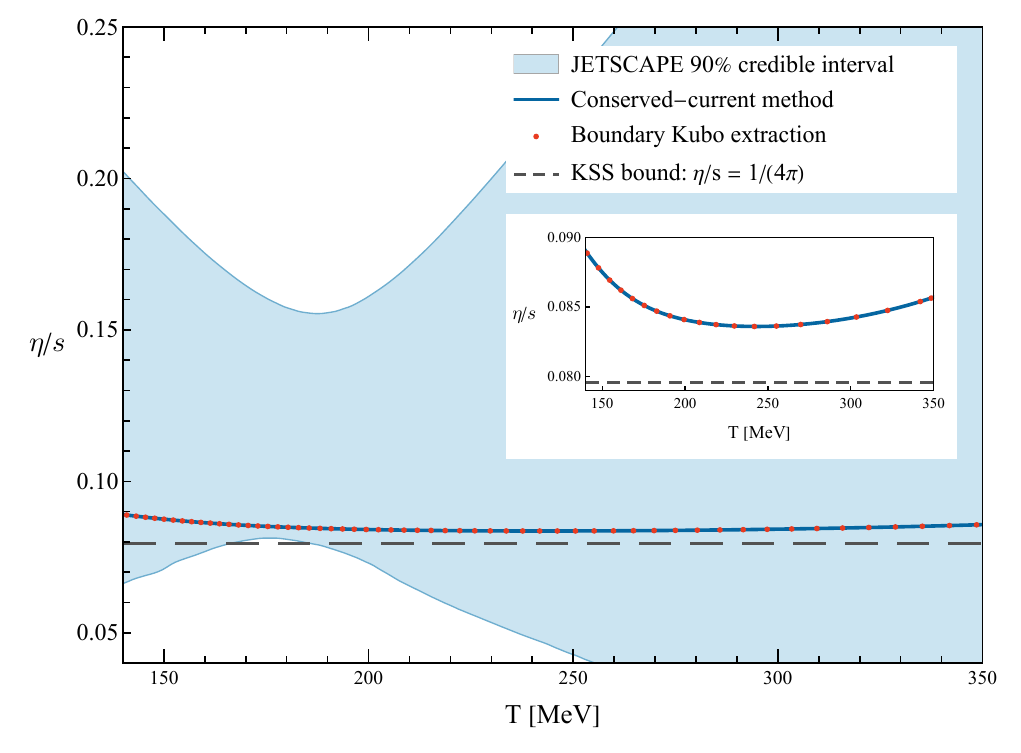}
    \caption{Shear-viscosity-to-entropy-density ratio $\eta/s$ as a function of temperature for the scalar-dependent coupling $H(\phi)=1-\phi^6/(1+2\phi^6)$ with $\alpha=-0.008$. Red circles: boundary Kubo result; blue solid curve: conserved-current result of Ref.~\cite{Tong:2026dwb}. The two results overlap within plotting accuracy. Both lie above the KSS bound $1/(4\pi)$ (black dashed line)~\cite{Kovtun:2003wp,Kovtun:2004de} and inside the JETSCAPE $90\%$ Bayesian credible interval~\cite{JETSCAPE:2020mzn} (blue shaded band). The inset shows a zoomed-in view of the curves and the KSS bound.
}
    \label{fig:eta_nonconstant_uv_horizon}
\vspace{4pt}

\centering
    \includegraphics[width=0.68\textwidth]{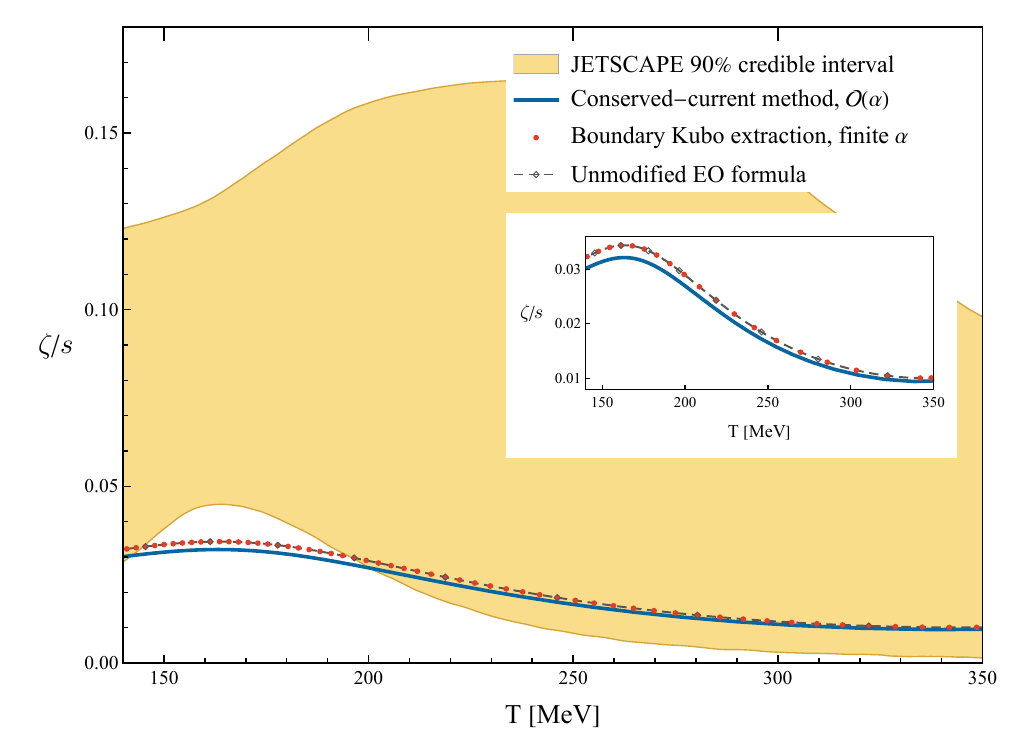}
    \caption{Bulk-viscosity-to-entropy-density ratio $\zeta/s$ as a function of temperature for the scalar-dependent coupling $H(\phi)=1-\phi^6/(1+2\phi^6)$ with $\alpha=-0.008$. Red circles: boundary Kubo result with full $\alpha$ dependence; gray dashed curve with open diamonds: unmodified EO formula, Eq.~\eqref{eq:EO-unmod}; blue solid curve: $\mathcal{O}(\alpha)$ conserved-current result of Ref.~\cite{Tong:2026dwb}. The boundary Kubo result agrees with the unmodified EO formula over the plotted temperature range. The conserved-current result has nearly the same peak temperature but lies visibly lower. The yellow shaded band shows the JETSCAPE $90\%$ Bayesian credible interval~\cite{JETSCAPE:2020mzn}. The inset shows a zoomed-in view of the curves.}
    \label{fig:zeta_nonconstant_uv_horizon}
\end{figure}

For the shear viscosity, Figure~\ref{fig:eta_nonconstant_uv_horizon} compares the boundary Kubo result~\eqref{eq:eta_over_s_uv} with the conserved-current result of Ref.~\cite{Tong:2026dwb}. The two agree graphically across the entire temperature range. Unlike the monotonic behavior in the constant-coupling case, $\eta/s$ develops a local minimum. Both curves remain above the KSS bound $1/(4\pi)$~\cite{Kovtun:2003wp,Kovtun:2004de}. Notably, they lie entirely within the JETSCAPE $90\%$ Bayesian credible interval~\cite{JETSCAPE:2020mzn} throughout the plotted range.

For the bulk viscosity, Figure~\ref{fig:zeta_nonconstant_uv_horizon} compares three independent approaches. The unmodified EO formula~\eqref{eq:EO-unmod} yields results almost identical to the boundary Kubo result across the entire temperature range. Both capture the prominent nonconformal peak near the crossover and the rapid falloff at high temperatures. In contrast, the $\mathcal{O}(\alpha)$ conserved-current result of Ref.~\cite{Tong:2026dwb} shares the same peak location and temperature trend, but is systematically lower. Phenomenologically, all curves lie near or within the JETSCAPE $90\%$ Bayesian credible interval~\cite{JETSCAPE:2020mzn}. They thus provide a reasonable order-of-magnitude estimate.

The boundary Kubo result agrees with the unmodified EO formula. To our knowledge, this is the first numerical verification of the formula in scalar-dependent Gauss--Bonnet gravity. 
Since our planar horizon obeys the area law, this result extends the conjecture of Ref.~\cite{Buchel:2024umq}, which links the validity of the unmodified EO formula to the area law of the horizon entropy, to a scalar-dependent coupling. The conjecture has previously been tested only for constant higher-curvature couplings.

We associate the gap between the conserved-current and boundary Kubo results for $\zeta/s$ with the approximation in Ref.~\cite{Tong:2026dwb}. As noted in Section~\ref{sec:introduction}, Ref.~\cite{Tong:2026dwb} truncated the calculation at linear order in $\alpha$. The visible discrepancy is consistent with the error introduced by this truncation. Moreover, even with the full $\alpha$ dependence, the bulk viscosity does not lie entirely within the JETSCAPE credible interval. This indicates that the present coupling function $H(\phi)$ needs further improvement.

\section{Conclusion and outlook}
\label{sec:conclusion}
We developed a boundary Kubo framework for computing shear and bulk viscosities in scalar-coupled Gauss--Bonnet gravity. The viscosities are extracted from the low-frequency limits of the renormalized retarded stress-tensor correlators. Curvature and scalar counterterms supplement those used for equilibrium thermodynamics and render the correlators finite, while the fluctuation equations are solved without expanding in the Gauss--Bonnet coupling. The resulting viscosities provide an independent benchmark for the horizon formulas.

We applied the framework to an EDGB model of holographic QCD, calibrated to lattice QCD thermodynamics. It arbitrates among the horizon prescriptions. For constant Gauss--Bonnet coupling, our results reproduce the conserved-current results for both viscosities. The bulk viscosity also agrees with that of the unmodified EO formula~\eqref{eq:EO-unmod}. For scalar-dependent coupling, the shear viscosity still agrees with the conserved-current result and develops a local minimum. The bulk viscosity agrees with that of the unmodified EO formula throughout the plotted range. In contrast, the conserved-current result for $\zeta/s$ in Ref.~\cite{Tong:2026dwb}, truncated at linear order in $\alpha$, lies visibly below our result. This difference reflects the error of the linear truncation. Phenomenologically, $\eta/s$ lies within the JETSCAPE credible interval~\cite{JETSCAPE:2020mzn}. However, $\zeta/s$ does not lie entirely within it.

These results show that the unmodified EO formula remains valid for the scalar-dependent coupling considered here. Since the horizon entropy still obeys the area law, this supports the conjecture of Ref.~\cite{Buchel:2024umq} beyond constant couplings. On the other hand, horizon methods bypass the retarded correlators and yield only their zero-frequency, zero-momentum limit. Accordingly, they cannot access information at finite frequency and momentum, such as spectral functions and the dispersion relations of hydrodynamic modes. Our framework has no such restriction. Inevitably, our study also has some limitations. We only examined a single coupling function $H(\phi)$. Moreover, the agreement with the unmodified EO formula is numerical rather than analytic.

Several directions follow from this work. The most pressing theoretical task is to derive the unmodified EO formula analytically in scalar-dependent Gauss--Bonnet gravity, including the role of the horizon area law and the conditions under which the prefactor $1/(4\pi)$ survives. On the phenomenological side, the framework can be extended to multiple scalar fields, finite baryon density, charged and anisotropic plasmas, and to finite frequency and momentum, where spectral functions and quasinormal modes provide additional constraints. Combining the finite-$\alpha$ results for both $\eta/s$ and $\zeta/s$ with Bayesian analyses of heavy-ion data would allow us to constrain the coupling function $H(\phi)$ and the model parameters simultaneously, rather than benchmark them against a single credible interval. Such a program would test the horizon prescriptions beyond the present setting and provide a more systematic bridge between holographic QCD and transport phenomenology.

\acknowledgments
We thank Chenwei Tong for providing the numerical data and for helpful discussions. This work was supported by the National Natural Science Foundation of China
(Grant Nos.~12525503, 12475053, 12588101, 12235016, and 12447101), by subproject funding from the Chinese Academy of Sciences for the Gravitational Redshift Measurement Scientific Experiment and Frontier Research
in Gravitational Physics, and by the Strategic Priority Research Program on Space Science of the Chinese Academy of Sciences (Grant Nos.~XDA30040000 and XDA30030000).

\appendix
\section{Equations of motion and asymptotic expansions}
\label{app:background}
\subsection{Background equations of motion}\label{sec:A.1}
Varying the action in Eq.~\eqref{eq2.1} with respect to $\phi$ and $g_{\mu\nu}$ yields
\begin{subequations}
\begin{align}
\nabla_\mu\nabla^\mu \phi
+\alpha \partial_\phi H(\phi)\mathcal{G}
-\partial_\phi V(\phi)
&=0,
\label{eq2}
\\[0.8em]
\begin{split}
\mathcal{R}_{\mu\nu}
-\frac{1}{2}g_{\mu\nu}\mathcal{R}
+\alpha H(\phi)L_{\mu\nu}
&+4\alpha P_{\mu\alpha\nu\beta}
\nabla^\alpha\nabla^\beta H(\phi)
\\
&=
\frac{1}{2}\nabla_\mu\phi\nabla_\nu\phi
-\frac{1}{2}g_{\mu\nu}
\left[
V(\phi)
+\frac{1}{2}\nabla_\lambda\phi\nabla^\lambda\phi
\right].
\end{split}
\label{eq3}
\end{align}
\end{subequations}
Here, $L_{\mu\nu}$ denotes the Lanczos tensor~\cite{Lanczos:1938sf,Lovelock:1971yv}, explicitly defined by
\begin{align}
    L_{\mu \nu}=-\frac{1}{2} g_{\mu \nu} \mathcal{G}+2 \mathcal{R}_{\mu \alpha \beta \gamma} P_\nu{}^{\alpha \beta \gamma},
\end{align}
where the tensor $P_{\mu\nu\rho\sigma}$ is given by
\begin{align}
    P_{\mu \nu \rho \sigma}=\mathcal{R}_{\mu \nu \rho \sigma}-2 g_{\mu[\rho} \mathcal{R}_{\sigma] \nu}+2 g_{\nu[\rho} \mathcal{R}_{\sigma] \mu}+g_{\mu[\rho} g_{\sigma] \nu} \mathcal{R}.
\end{align}

Substituting the metric ansatz~\eqref{eq2.2} into Eqs.~\eqref{eq2} and~\eqref{eq3}, we obtain the following equations:

\begin{subequations}
\begin{align}
&
\phi^{\prime\prime}
+\left(
\frac{3}{r}
-\frac{\eta^{\prime}}{2}
+\frac{f^{\prime}}{f}
\right)\phi^{\prime}
+\frac{1}{f}\bigg[
\frac{12\alpha f f^{\prime\prime}H_\phi}{r^2}
+\frac{24\alpha f f^{\prime}H_\phi}{r^3}
-\frac{30\alpha f f^{\prime}\eta^{\prime}H_\phi}{r^2}
\nonumber\\
&\hspace{1.8cm}
+\frac{12\alpha f^{\prime 2}H_\phi}{r^2}
-\frac{12\alpha f^2\eta^{\prime}H_\phi}{r^3}
-\frac{12\alpha f^2\eta^{\prime\prime}H_\phi}{r^2}
+\frac{6\alpha f^2\eta^{\prime 2}H_\phi}{r^2}
-\partial_{\phi}V(\phi)
\bigg]
=0 ,
\label{eq7}
\\[1.1em]
&
\frac{\eta^{\prime}}{r}
+\frac{1}{3}\phi^{\prime 2}
-\frac{4\alpha f H(\phi)\eta^{\prime}}{r^3}
-\frac{12\alpha f H_\phi\eta^{\prime}\phi^{\prime}}{r^2}
-\frac{8\alpha f \phi^{\prime 2}H_{\phi\phi}}{r^2}
-\frac{8\alpha f H_\phi\phi^{\prime\prime}}{r^2}
=0 ,
\label{eq8}
\\[1.1em]
&
\frac{2V(\phi)}{3f}
+\frac{4}{r^2}
+\frac{2f^{\prime}}{fr}
-\frac{\eta^{\prime}}{r}
-\frac{8\alpha H(\phi)f^{\prime}}{r^3}
+\frac{4\alpha fH(\phi)\eta^{\prime}}{r^3}
-\frac{16\alpha fH_\phi\phi^{\prime}}{r^3}
\nonumber\\
&\hspace{1.8cm}
-\frac{24\alpha f^{\prime}H_\phi\phi^{\prime}}{r^2}
+\frac{12\alpha fH_\phi\eta^{\prime}\phi^{\prime}}{r^2}
-\frac{8\alpha f\phi^{\prime 2}H_{\phi\phi}}{r^2}
-\frac{8\alpha fH_\phi\phi^{\prime\prime}}{r^2}
=0 .
\label{eq9}
\end{align}
\end{subequations}
Here, primes denote radial derivatives, while subscripts denote scalar derivatives, with $H_\phi \equiv \partial_\phi H$ and $H_{\phi\phi} \equiv \partial_\phi^2 H$.

\subsection{IR and UV expansions}
\label{app:IR_UV_expansions}
To obtain numerical solutions, we impose regularity at the event horizon $r=r_h$ (the IR), and require the geometry to approach $\mathrm{AdS}_5$ as $r\to\infty$ (the UV). Horizon regularity gives the Taylor expansions:
\begin{align}\label{eq11}
    \begin{aligned}
f & =f_h^1\left(r-r_h\right)+\cdots, \\[4pt]
\eta & =\eta_h+\eta_h^1\left(r-r_h\right)+\cdots, \\[4pt]
\phi & =\phi_h+\phi_h^1\left(r-r_h\right)+\cdots .
\end{aligned}
\end{align}
Matching Eq.~\eqref{eq11} to Eqs.~\eqref{eq7}--\eqref{eq9} order by order leaves three independent IR parameters, $\left(r_h,\eta_h,\phi_h\right)$, which determine all remaining coefficients. For numerical convenience, we shall use the scaling symmetry to set $r_h=1$ and choose $\eta_h=0$. The entire family of solutions is then parameterized solely by $\phi_h$.

Near the asymptotically AdS boundary, the corresponding UV expansions are
\begingroup
\allowdisplaybreaks[2]
\begin{subequations}\label{eq:bg_uv}
\begin{align}
\phi(r)
={}&
\frac{\phi_s}{r}
-\frac{3c_2\phi_s^2}{r^2}
+\frac{\phi_v}{r^3}
-\frac{
\left[1-(1-4\alpha)(54c_2^2+6c_1^4)\right]\phi_s^3
}{
6(1-4\alpha)r^3
}\ln r
+\mathcal{O}\!\left(\frac{\ln r}{r^4}\right),
\label{eq:phi_uv}
\\[0.8em]
f(r)
={}&
r^2
+\frac{\phi_s^2}{6(1-4\alpha)}
-\frac{4c_2\phi_s^3}{3(1-4\alpha)r}
+\frac{f_v}{r^2}
-\frac{
\left[1-(1-4\alpha)(54c_2^2+6c_1^4)\right]\phi_s^4
}{
12(1-4\alpha)^2r^2
}\ln r
\nonumber\\
&+\mathcal{O}\!\left(\frac{\ln r}{r^3}\right),
\label{eq:f_uv}
\\[0.8em]
\eta(r)
={}&
\frac{\phi_s^2}{6(1-4\alpha)r^2}
-\frac{4c_2\phi_s^3}{3(1-4\alpha)r^3}
+\frac{\phi_s\phi_v}{2(1-4\alpha)r^4}
+\frac{
\left[1+4\alpha+(1-4\alpha)^2(378c_2^2-6c_1^4)\right]\phi_s^4
}{
144(1-4\alpha)^3r^4
}
\nonumber\\
&-\frac{
\left[1-(1-4\alpha)(54c_2^2+6c_1^4)\right]\phi_s^4
}{
12(1-4\alpha)^2r^4
}\ln r
+\mathcal{O}\!\left(\frac{\ln r}{r^5}\right).
\label{eq:eta_uv}
\end{align}
\end{subequations}
\endgroup
Here, we have normalized the boundary time coordinate such that $\eta(r\to\infty)=0$, for which the Hawking temperature matches the boundary field theory temperature.

\section{Holographic renormalization}\label{sec:B1}
In this appendix, we specify the boundary action and counterterms summarized in Section~\ref{Counterterms} for the EDGB model. Following standard holographic renormalization~\cite{deHaro:2000vlm,Skenderis:2002wp,Anastasiou:2025usa,Balasubramanian:1999re}, the total renormalized action is defined as
\begin{align}
S_{\rm ren}
=S_{\rm bulk}+S_{\partial},\quad
S_{\partial}=S_{\rm GHY}+S_{\rm Myers}+S_{\rm ct},
\end{align}
where $S_{\rm bulk}$ is the bulk action given in Eq.~\eqref{eq2.1}, while
\begin{align}
S_{\rm GHY}
={}&
\frac{1}{2\kappa_5^2}
\int_{\partial\mathcal{M}} \dd^4x \sqrt{-\gamma}\,2K,
\label{eq:SGHY_hr}
\\[0.6em]
S_{\rm Myers}
={}&
\frac{1}{2\kappa_5^2}
\int_{\partial\mathcal{M}} \dd^4x \sqrt{-\gamma}\,
4\alpha H(\phi)
\left(
J-2\widehat G_{\mu\nu}K^{\mu\nu}
\right)
\label{eq:SMyers_hr}
\end{align}
are, respectively, the Gibbons--Hawking--York boundary term~\cite{Gibbons:1976ue} and the Myers boundary term~\cite{Myers:1987yn,Deruelle:2018vtt}. They ensure a well-posed variational principle. We take $n^\mu$ to be the outward-pointing unit normal to $\partial\mathcal M$. Then $K_{\mu\nu}=\gamma_\mu{}^\rho\nabla_\rho n_\nu$ is the extrinsic curvature with trace $K=\gamma^{\mu\nu}K_{\mu\nu}$, and $\widehat G_{\mu\nu}=\widehat R_{\mu\nu}-\tfrac12\gamma_{\mu\nu}\widehat R$ is the boundary Einstein tensor. The tensor $J_{\mu\nu}$ is given by~\cite{Davis:2002gn}
\begin{align}
J_{\mu\nu}
=
\frac{1}{3}
\left(
2K K_{\mu\rho}K^\rho{}_\nu
+K_{\rho\sigma}K^{\rho\sigma}K_{\mu\nu}
-2K_{\mu\rho}K^{\rho\sigma}K_{\sigma\nu}
-K^2K_{\mu\nu}
\right),
\qquad
J=J^\mu{}_\mu .
\label{eq:Jmunu_def}
\end{align}

As discussed in Section~\ref{Counterterms}, $S_{\rm ct}$ consists of two parts. The intrinsic-curvature and scalar terms introduced in this work are given in Eq.~\eqref{eq:Sct_summary}. The established part from Refs.~\cite{Tong:2025rxz,Tong:2026dwb,Cai:2022omk}, denoted by $S_{\rm ct}^{(0)}$, reads

\begin{align}
S_{\rm ct}^{(0)}
=
-\frac{1}{2\kappa_5^2}
\int_{\partial\mathcal{M}} \dd^4x\sqrt{-\gamma}\,
\bigg[
&6-8\alpha
+\frac{1}{2}\phi^2
-c_2\phi^3
+b\phi^4
\nonumber\\
&+\left(\frac{1}{2}c_1^4+\frac{9}{2}c_2^2+\frac{1}{12(4\alpha-1)}\right)\phi^4\ln r
\bigg],
\label{eq:Sct_flat_def}
\end{align}
where the finite coefficient $b$ fixes the renormalization scheme and does not affect either viscosity.

Combining the above contributions, the boundary action $S_{\partial}$ takes the form
\begin{align}
S_{\partial}
={}&
\frac{1}{2\kappa_5^2}
\int_{\partial\mathcal{M}} \dd^4x\sqrt{-\gamma}
\bigg\{
2K
+4\alpha H(\phi)
\left(
J-2\widehat G_{\mu\nu}K^{\mu\nu}
\right)
\nonumber\\
&\hspace{1.2cm}
-\bigg[
6-8\alpha
+\frac{1}{2}\phi^2
-c_2\phi^3
+b\phi^4
+\left(
\frac{1}{2}c_1^4
+\frac{9}{2}c_2^2
+\frac{1}{12(4\alpha-1)}
\right)\phi^4\ln r
\nonumber\\
&\hspace{1.2cm}
+\left(\frac{1}{2}+2\alpha\right)\widehat R
+\left(\frac{1}{4}-\alpha\right)
\left(
\widehat R_{\mu\nu}\widehat R^{\mu\nu}
-\frac{1}{3}\widehat R^2
\right)\ln r
+\frac{1}{2}
\left(
\phi\widehat\Box\phi
-\frac{1}{6}\widehat R\phi^2
\right)\ln r
\bigg]
\bigg\}.
\label{eq:Sboundary_general_H}
\end{align}

The Brown--York stress tensor is then defined as
\begin{align}
T_{\mu\nu}
&=
\lim_{r\to\infty}
\left(
-\frac{2r^2}{\sqrt{-\gamma}}
\frac{\delta S_{\rm ren}^{\mathrm{on\text{-}shell}}}
{\delta\gamma^{\mu\nu}}
\right)
\nonumber\\
&=
-\frac{1}{2\kappa_5^2}
\lim_{r\to\infty}r^2
\bigg\{
2(K_{\mu\nu}-K\gamma_{\mu\nu})
+(6-8\alpha)\gamma_{\mu\nu}
+4\alpha H(\phi)
\left(
3J_{\mu\nu}
-J\gamma_{\mu\nu}
+2\widehat P_{\mu\rho\sigma\nu}K^{\rho\sigma}
\right)
\nonumber\\
&\hspace{0.8cm}
+2\Pi_{\mu\nu}
+\gamma_{\mu\nu}
\left[
\frac{1}{2}\phi^2
-c_2\phi^3
+b\phi^4
+\left(
\frac{1}{2}c_1^4
+\frac{9}{2}c_2^2
+\frac{1}{12(4\alpha-1)}
\right)\phi^4\ln r
\right]
\nonumber\\
&\hspace{0.8cm}
-(1+4\alpha)\widehat G_{\mu\nu}
+\left(\frac{1}{2}-2\alpha\right)
\widehat{\mathcal B}_{\mu\nu}\ln r
+\frac{1}{6}\ln r
\left(
\phi^2\widehat G_{\mu\nu}
+\gamma_{\mu\nu}\widehat\Box\phi^2
-\widehat\nabla_\mu\widehat\nabla_\nu\phi^2
\right)
\nonumber\\
&\hspace{0.8cm}
+\ln r
\left(
\widehat\nabla_\mu\phi\,\widehat\nabla_\nu\phi
-\frac{1}{2}\gamma_{\mu\nu}
\widehat\nabla_\rho\phi\,\widehat\nabla^\rho\phi
\right)
\bigg\},
\label{eq:Tmunu_general_H}
\end{align}
where $\Pi_{\mu\nu}$ collects the normal and tangential derivatives of $H(\phi)$~\cite{Julie:2020vov}:
\begin{align}
\Pi_{\mu\nu}
={}&
4\alpha
\left[n^\lambda\nabla_\lambda H(\phi)\right]
\bigg[
\widehat G_{\mu\nu}
+\frac{1}{2}\gamma_{\mu\nu}
\left(
K^2-K_{\rho\sigma}K^{\rho\sigma}
\right)
-KK_{\mu\nu}
+K_{\mu\rho}K^\rho{}_\nu
\bigg]
\nonumber\\
&+
4\alpha
\bigg[
\gamma_{\mu\nu}
\left(
K\widehat\Box H(\phi)
-K^{\rho\sigma}
\widehat\nabla_\rho\widehat\nabla_\sigma H(\phi)
\right)
-K_{\mu\nu}\widehat\Box H(\phi)
-K\widehat\nabla_\mu\widehat\nabla_\nu H(\phi)
\nonumber\\
&\hspace{1.2cm}
+K_\mu{}^\rho
\widehat\nabla_\rho\widehat\nabla_\nu H(\phi)
+K_\nu{}^\rho
\widehat\nabla_\rho\widehat\nabla_\mu H(\phi)
\bigg],
\label{eq:Pimunu_def_general_H}
\end{align}
and the Bach tensor $\widehat{\mathcal{B}}_{\mu\nu}$~\cite{Gorbatenko:2001fx} is given explicitly by
\begin{align}
\widehat{\mathcal{B}}_{\mu\nu}
={}&
-2\widehat R_{\mu\rho\nu\sigma}\widehat R^{\rho\sigma}
+\frac{2}{3}\widehat R\,\widehat R_{\mu\nu}
+\frac{1}{2}\gamma_{\mu\nu}
\left(\widehat R_{\rho\sigma}\widehat R^{\rho\sigma}-\frac{1}{3}\widehat R^2\right)
\nonumber\\
&
-\widehat\Box\widehat R_{\mu\nu}
+\frac{1}{3}\widehat\nabla_\mu\widehat\nabla_\nu\widehat R
+\frac{1}{6}\gamma_{\mu\nu}\widehat\Box\widehat R .
\label{eq:Bhat_def_general_H}
\end{align}

Since $H(\phi)=1+\mathcal O(\phi^6)$, the $H$-dependent terms, $\Pi_{\mu\nu}$ and the $(H-1)$ part of the Myers term, introduce no additional divergences. The boundary action~\eqref{eq:Sboundary_general_H} thus requires no $H(\phi)$-dependent counterterms. A counterterm action for general $H(\phi)$ is beyond the present scope.

\section{Details of the viscosity calculations}\label{sec:C}
\subsection{Shear viscosity}
\label{sec:C_shear}
We give the explicit coefficients of Eq.~\eqref{eq:shear_master} and derive the boundary response used in Eq.~\eqref{eq:eta_over_s_uv}. For the zero-momentum tensor mode, the quadratic effective action can be written as~\cite{Cai:2009zv}
\begin{align}
S_{\rm eff}=-\frac{1}{4\kappa_5^2}\int\dd^5x\left[A(\partial_r h)^2-B(\partial_t h)^2\right],
\label{eq:shear_quadratic_action}
\end{align}
up to boundary terms. Here $h = h(r)e^{-i\omega t}$, and the coefficients $A$ and $B$ are given by
\begin{align}
A=r^3f e^{-\eta/2}K_r,\qquad B=\frac{r^3e^{\eta/2}}{f}K_t,
\label{eq:shear_kinetic_coefficients}
\end{align}
where the Gauss--Bonnet correction factors read
\begin{subequations}\label{eq:shear_K_factors}
\begin{align}
K_r &= 1-2\alpha\left[\left(\frac{H}{r}+H_r\right)(f'-f\eta')+\frac{2fH_r}{r}\right],
\label{eq:shear_Kr}
\\[4pt]
K_t &= 1-2\alpha\left[\left(\frac{H}{r}+H_r\right)f'+\frac{2fH_r}{r}+2fH_{rr}\right].
\label{eq:shear_Kt}
\end{align}
\end{subequations}
We abbreviate $H=H(\phi(r))$, $H_r=H_\phi\phi'$, and $H_{rr}=H_{\phi\phi}(\phi')^2+H_\phi\phi''$, with primes denoting radial derivatives. Varying Eq.~\eqref{eq:shear_quadratic_action} with respect to $h$ and dividing it by $A$ then gives the linearized equation~\eqref{eq:shear_master}, with the explicit coefficients
\begin{align}
\mathcal P_{\mathrm{GB}}
={}&\frac{K_r'}{K_r}
=-\frac{2\alpha}{K_r}\bigg[
\left(H_{rr}+\frac{H_r}{r}-\frac{H}{r^2}\right)(f'-f\eta')
+\left(\frac{H}{r}+H_r\right)(f''-f'\eta'-f\eta'')
\nonumber\\
&\hphantom{\frac{K_r'}{K_r}={}}{+}\frac{2f'H_r}{r}+\frac{2fH_{rr}}{r}-\frac{2fH_r}{r^2}
\bigg],
\label{eq:shear_P_GB}
\\
\mathcal Q_{\mathrm{GB}}
={}&\frac{e^\eta}{f^2}\left(\frac{K_t}{K_r}-1\right)
=-\frac{2\alpha e^\eta}{fK_r}
\left[\left(\frac{H}{r}+H_r\right)\eta'+2H_{rr}\right].
\label{eq:shear_Q_GB}
\end{align}

To select the retarded solution, we impose the infalling condition at the horizon $r_h$~\cite{Son:2002sd,Iqbal:2008by},
\begin{align}
h(r)\sim(r-r_h)^{-i\omega/(4\pi T)},\quad r\to r_h.
\end{align}
Inserting the UV expansions of $h(r)$ and of the background into $T_{xy}$, the $xy$ component of Eq.~\eqref{eq:Tmunu_general_H}, we obtain the finite response
\begin{align}
\delta\left\langle T_{xy}(\omega)\right\rangle=\frac{1}{2\kappa_5^2} \left[4(1-4\alpha) h_v-\mathcal{C}_{xy} h_s\right],
\end{align}
where the contact coefficient
\begin{align}
\mathcal{C}_{xy}
={}&(1-4\alpha)f_v-\phi_s\phi_v
-\frac{\phi_s^2\omega^2}{16}
+\frac{3}{16}(1-4\alpha)\omega^4
+b\phi_s^4
\nonumber\\
&+\phi_s^4\left(
3c_2^2+\frac{c_1^4}{6}
+\frac{28\alpha-9}{144(1-4\alpha)^2}
\right)
\end{align}
is purely real. On the other hand, linear response theory~\cite{Kovtun:2012rj},
with the convention in Eq.~\eqref{eq:retarded_convention}, gives
\begin{align}
\delta\left\langle T_{xy}(\omega)\right\rangle
=-G_{\eta}^R(\omega, \mathbf{k}=\mathbf{0})h_s,
\end{align}
and hence
\begin{align}
G_{\eta}^R(\omega, \mathbf{k}=\mathbf{0})=-\dfrac{\delta\left\langle T_{x y}(\omega)\right\rangle}{h_s}=\frac{1}{2 \kappa_5^2}\left[-4(1-4 \alpha) \frac{h_{v}}{h_s}+\mathcal{C}_{xy}\right].
\end{align}
The Kubo formula then gives the shear viscosity
\begin{align}
\eta=-\lim_{\omega\to 0}\frac{1}{\omega}\mathrm{Im}\,G_{\eta}^R(\omega, \mathbf{k}=\mathbf{0}) =\frac{4(1-4\alpha)}{2 \kappa_5^2}\lim_{\omega\to 0}\frac{1}{\omega}\mathrm{Im}\,\left(\frac{h_{v}}{h_s}\right).
\end{align} 
Finally, dividing by the entropy density immediately reproduces the ratio $\eta/s$ presented in Eq.~\eqref{eq:eta_over_s_uv}.

\subsection{Bulk viscosity}
\label{sec:C_bulk}
We first verify the gauge invariance of $Z(r)$ defined in Eq.~\eqref{eq:Zphi_definition} and present the explicit coefficients of the master equation~\eqref{eq:bulk_master}. Finally, we evaluate the boundary response to derive the ratio $\zeta/s$ in Eq.~\eqref{eq:zeta_over_s_uv}.

In the zero-momentum, isotropic sector, we consider an infinitesimal
gauge transformation generated by
\begin{align}
\xi^\mu
=
\bigl(\xi^t(t,r),\,\xi^r(t,r),\,\mathbf{0}\bigr),
\qquad \mu=t,r,x,y,z.
\label{eq:bulk_xi_parameter}
\end{align}
At linear order, the scalar and metric perturbations transform as
\begin{subequations}\label{eq:bulk_lie_transformations}
\begin{align}
\delta\phi
&\longrightarrow
\delta\phi+\mathcal{L}_\xi\phi
=
\delta\phi+\xi^\mu\partial_\mu\phi,
\label{eq:bulk_lie_phi}
\\[0.4em]
\delta g_{\mu\nu}
&\longrightarrow
\delta g_{\mu\nu}+(\mathcal{L}_\xi g)_{\mu\nu}
\nonumber\\
&=
\delta g_{\mu\nu}
+\xi^\rho\partial_\rho g_{\mu\nu}
+g_{\rho\nu}\partial_\mu\xi^\rho
+g_{\mu\rho}\partial_\nu\xi^\rho,
\label{eq:bulk_lie_metric}
\end{align}
\end{subequations}
where $\mathcal{L}_\xi$ denotes the Lie derivative along $\xi^\mu$. Since the background is static and $\partial_i\xi^\mu=0$, the temporal component $\xi^t$ does not contribute to the scalar and spatial metric variations. With $\phi=\phi(r)$ and $g_{ii}=r^2$, the perturbations transform as
\begin{subequations}\label{eq:bulk_gauge_variations}
\begin{align}
\delta\phi
&\longrightarrow
\delta\phi+\xi^r\phi'(r),
\label{eq:bulk_gauge_var_phi}
\\[0.3em]
\delta g_{ii}
&\longrightarrow
\delta g_{ii}+2r\xi^r,
\label{eq:bulk_gauge_var_gii}
\end{align}
\end{subequations}
with no sum over $i=x,y,z$. Recalling the perturbation $\delta g_{ii}=r^2h_{11}$, it follows that
\begin{align}
h_{11}
\longrightarrow
h_{11}+\frac{2}{r}\xi^r.
\label{eq:bulk_delta_h11}
\end{align}
Consequently, $Z=\delta\phi-\frac{r\phi'(r)}{2}h_{11}$ satisfies
\begin{align}
Z \longrightarrow
\delta\phi+\xi^r\phi'(r)
-\frac{r\phi'(r)}{2}
\left(h_{11}+\frac{2}{r}\xi^r\right)
= Z,
\label{eq:bulk_gauge_invariance}
\end{align}
which is gauge invariant at linear order.

Rather than expanding the action to quadratic order, we work directly with the linearized equations of motion. Eliminating $h_{00}(r)$ and $h_{11}(r)$ via the linearized Einstein equations decouples the system into the master equation~\eqref{eq:bulk_master}. This confirms that $Z(r)$ captures the single propagating degree of freedom in this channel. We define the auxiliary background combinations
\begin{align}
\Delta_Z&=1-\frac{4\alpha fH}{r^2}
-\frac{12\alpha fH_\phi\phi'}{r},
\nonumber\\
\Theta_Z&=\frac{1}{\Delta_Z}\left\{
\frac{r\phi'}{3}+\frac{4\alpha f}{r}
\left[H_\phi\left(\frac{f'}{f}-\eta'\right)-2H_{\phi\phi}\phi'\right]\right\}.
\label{eq:bulk_common_factors}
\end{align}
In terms of these quantities, the factors $K_{Z,r}$ and $K_{Z,t}$ entering the bulk master equation~\eqref{eq:bulk_master} are expressed as
\begin{subequations}\label{eq:bulk_kinetic_potential}
\begin{align}
K_{Z,r}={}&1+\frac{8\alpha fH_\phi\phi'}{r\Delta_Z}
+\frac{96\alpha^2f^2H_\phi^2}{r^3\Delta_Z}
\left[3\left(\frac{f'}{f}-\eta'\right)+\frac{2}{r}\right]
\nonumber\\
&+\frac{64\alpha^2fH_\phi^2}{r^2\Delta_Z^2}
\bigg\{\frac{V}{2}+\frac{6\alpha f^2}{r^3}
\bigg[H\left(\frac{f'}{f}-\eta'\right)
+\left(3r\left(\frac{f'}{f}-\eta'\right)+2\right)H_\phi\phi'\bigg]\bigg\},
\label{eq:bulk_radial_kinetic}
\\[0.4em]
K_{Z,t}={}&1+\frac{8\alpha fH_\phi\phi'}{r\Delta_Z}
+\frac{48\alpha^2f^2H_\phi^2}{r^3\Delta_Z}
\left[\frac{4f'}{f}-\eta'-\frac{2\Delta_Z'}{\Delta_Z}\right],
\label{eq:bulk_temporal_kinetic}
\end{align}
\end{subequations}
and $\mathcal{V}_Z$ is given by
\begin{align}
\mathcal V_Z={}&V_{\phi\phi}-\alpha H_{\phi\phi}\mathcal G
+\frac{V}{2} \Theta_Z^2+V_\phi \Theta_Z
\nonumber\\
&+\frac{6\alpha f^2 \Theta_Z}{r^3}\bigg\{
\left(H \Theta_Z-2H_\phi\right)\left(\frac{f'}{f}-\eta'\right)
+\left[3r\left(\frac{f'}{f}-\eta'\right)+2\right]
\left(H_\phi \Theta_Z-2H_{\phi\phi}\right)\phi'\bigg\}
\nonumber\\
&+\frac{e^{\eta/2}}{r^3}\frac{\dd}{\dd r}\Bigg\{r^3fe^{-\eta/2}\Bigg[
\left(\frac{\phi'}{2}+\frac{6\alpha fH_\phi}{r^2}
\left[3\left(\frac{f'}{f}-\eta'\right)+\frac{2}{r}\right]\right) \Theta_Z
+\frac{4\alpha H_\phi}{r\Delta_Z}\left(V \Theta_Z+V_\phi\right)
\nonumber\\
&\quad+\frac{48\alpha^2f^2H_\phi}{r^4\Delta_Z}\bigg\{
\left(H \Theta_Z-H_\phi\right)\left(\frac{f'}{f}-\eta'\right)
+\left[3r\left(\frac{f'}{f}-\eta'\right)+2\right]
\left(H_\phi \Theta_Z-H_{\phi\phi}\right)\phi'\bigg\}\Bigg]\Bigg\}.
\label{eq:bulk_effective_potential}
\end{align}
All primes denote radial derivatives, and subscripts $\phi$ denote scalar derivatives.
Crucially, these expressions retain the full, non-perturbative dependence on the Gauss--Bonnet coupling without performing any linear expansion in $\alpha$. 

As in the shear channel, $Z(r)$ satisfies the infalling boundary condition at the horizon:
\begin{align}   
Z(r)
\sim
(r-r_h)^{-i\omega/(4\pi T)}
.
\label{eq:bulk_ingoing_expansion}
\end{align}
In the UV region, the expansion of $Z(r)$ is given in Eq.~\eqref{eq:Zphi_uv_bulk}, while $h_{00}$ and $h_{11}$ admit the expansions
\begingroup
\small
\setlength{\jot}{3pt}
\allowdisplaybreaks[2]
\begin{subequations}\label{eq:bulk_uv}
\begin{align}
h_{11}(r)
={}&
h_{11s}
+\frac{\phi_s\left(h_{11s}\phi_s-2Z_s\right)}
{12(1-4\alpha)r^2}
+\frac{2c_2\phi_s^2\left(2Z_s-h_{11s}\phi_s\right)}
{3(1-4\alpha)r^3}
+\mathcal{O}\!\left(\frac{\ln r}{r^4}\right),
\label{eq:h11_uv_bulk}
\\[0.8em]
h_{00}(r)
={}&
h_{00s}
+\frac{h_{11s}\phi_s^2-2Z_s\phi_s+6(1-4\alpha)h_{11s}\omega^2}{12(1-4\alpha)r^2}
+\frac{2c_2\phi_s^2\left(2Z_s-h_{11s}\phi_s\right)}{3(1-4\alpha)r^3}
+\mathcal{O}\!\left(\frac{\ln r}{r^4}\right),
\label{eq:h00_uv_bulk}
\end{align}
\end{subequations}
\endgroup
where $h_{11s}$ and $h_{00s}$ denote the boundary metric sources.

Inverting Eq.~\eqref{eq:Zphi_definition}, the scalar fluctuation is reconstructed as
\begin{align}
\delta\phi(r) = Z(r) + \frac{r\phi'(r)}{2}h_{11}(r).
\label{eq:bulk_scalar_reconstruction}
\end{align}
At leading order in $1/r$, the UV expansions give
$\delta\phi_s = Z_s - \tfrac{\phi_s}{2}h_{11s}$, where $\delta\phi_s$ denotes the
scalar-source perturbation. Since the retarded correlator $G^R_\zeta$ is the response to the metric source at fixed scalar source, we set $\delta\phi_s=0$, which gives
\begin{align}
Z_s = \frac{\phi_s}{2}h_{11s}.
\label{eq:Zs_metric_source}
\end{align}

The correlator $G^R_\zeta(\omega,\mathbf{k}=\mathbf{0})$ in Eq.~\eqref{eq:kubo_zeta}
follows from the spatial trace $\mathcal T$, built from the spatial diagonal components of Eq.~\eqref{eq:Tmunu_general_H}. These components are equal by $\mathrm{SO}(3)$ isotropy. We insert the UV expansions of the background and of $h_{00}$, $h_{11}$, and $Z$ into these components. Linearizing in the perturbations yields the finite response

\begin{align}
\delta\langle T_{xx}(\omega)\rangle
&= \delta\langle T_{yy}(\omega)\rangle
 = \delta\langle T_{zz}(\omega)\rangle
\nonumber\\
&= -\frac{1}{2\kappa_5^2}
   \left(
     h_{11s}\mathcal{C}_{xx}
     - \frac{2}{3}\phi_s Z_v
   \right)
 = \frac{h_{11s}}{2\kappa_5^2}
   \left(
     \frac{\phi_s^2}{3}\frac{Z_v}{Z_s}
     - \mathcal{C}_{xx}
   \right),
\label{eq:bulk_stress_response}
\end{align}
where the last equality uses Eq.~\eqref{eq:Zs_metric_source}. The contact term $\mathcal{C}_{xx}$ is given by
\begin{align}
\mathcal{C}_{xx}
&= -(1-4\alpha)f_v + \phi_s\phi_v
\nonumber\\[4pt]
&\quad
+ \frac{\phi_s^4}{144(1-4\alpha)^2}
\Bigl[
5 - 12\alpha
+ (1-4\alpha)^2
  \bigl(144b + 756c_2^2 - 36c_1^4\bigr)
\Bigr].
\label{eq:Cxx_contact_term}
\end{align}
Being real and local, it does not contribute to the imaginary part in the Kubo formula. Linear response theory then gives, with the convention in Eq.~\eqref{eq:retarded_convention},\footnote{The factor $1/2$ arises because the metric source couples to the stress tensor through $\frac12 h_{\mu\nu}T^{\mu\nu}$. For the isotropic source $h_{11s}$, this coupling becomes $\frac12 h_{11s}\mathcal{T}$, so the source conjugate to $\mathcal{T}$ is $\frac12 h_{11s}$.
}

\begin{align}
\delta\langle \mathcal{T}(\omega)\rangle
&= -\frac{1}{2} G^R_{\zeta}(\omega,\mathbf{k}=\mathbf{0})\,h_{11s}.
\label{eq:linear_response_bulk}
\end{align}
Using $\delta\langle\mathcal{T}\rangle =\delta\langle T_{xx}\rangle +\delta\langle T_{yy}\rangle +\delta\langle T_{zz}\rangle=3\,\delta\langle T_{xx}\rangle$ and
Eq.~\eqref{eq:bulk_stress_response}, we obtain
\begin{align}
G^R_{\zeta}(\omega,\mathbf{k}=\mathbf{0})
= -\frac{2\,\delta\langle \mathcal{T}(\omega)\rangle}{h_{11s}}
=\frac{3}{\kappa_5^2}\mathcal{C}_{xx}
   - \frac{\phi_s^2}{\kappa_5^2}\frac{Z_v}{Z_s}.
\label{eq:GR_bulk_final}
\end{align}
The Kubo formula~\eqref{eq:kubo_zeta} then yields the bulk viscosity
\begin{align}
\zeta
&= -\frac{1}{9}\lim_{\omega\to0}\frac{1}{\omega}\mathrm{Im}\,G^R_{\zeta}(\omega,\mathbf{k}=\mathbf{0})
= \frac{\phi_s^2}{9\kappa_5^2}\lim_{\omega\to0}\frac{1}{\omega}\mathrm{Im}\left(\frac{Z_v}{Z_s}\right).
\end{align}
Combining this with the entropy density, we obtain the ratio $\zeta/s$ reported in Eq.~\eqref{eq:zeta_over_s_uv} straightforwardly.

\section{Details of the numerical calculations}
\label{app:numerical_details}
This appendix collects technical details of our numerical calculations. In Section~\ref{app:eo_numerical}, we rewrite the unmodified EO formula for numerical evaluation. In Section~\ref{app:omega_convergence}, we test the convergence of the Kubo extraction at low frequency.

\subsection{Numerical implementation of the EO formula}
\label{app:eo_numerical}

The unmodified EO formula is defined in Eq.~\eqref{eq:EO-unmod}. We numerically solve the equations at fixed $r_h=1$. The entropy density is the same for all scalar horizon values $\phi_h$, so we cannot apply Eq.~\eqref{eq:EO-unmod} directly. Meanwhile, the boundary source $\phi_s$ varies with $\phi_h$. The scaling symmetry in Appendix~\ref{app:IR_UV_expansions} lets us set the source of all solutions to a common value, $\phi_s\to\Phi_s=\lambda\phi_s$. Under this rescaling, the horizon radius becomes $\lambda r_h=\Phi_s/\phi_s$. The entropy density~\eqref{eq:entropy_density} then reads
\begin{equation}
s=\frac{2\pi}{\kappa_5^2}\left(\frac{\Phi_s}{\phi_s}\right)^3 .
\label{eq:s_physical_scaling}
\end{equation}
Since $\Phi_s$ and $\kappa_5$ are fixed across the thermal family, $\partial\ln s/\partial\ln\phi_s=-3$, and hence
\begin{equation}
\frac{\partial\phi_h}{\partial\ln s}=-\frac{1}{3}\frac{\partial\phi_h}{\partial\ln\phi_s}.
\label{eq:dphih_dlns}
\end{equation}

Substituting this into Eq.~\eqref{eq:EO-unmod} gives the working expression used in this work,
\begin{equation}
\left(\frac{\zeta}{s}\right)_{\mathrm{EO}}
=\frac{1}{36\pi}\left(\frac{\partial\phi_h}{\partial\ln\phi_s}\right)^2 .
\label{eq:EO_numerical_working}
\end{equation}
Both $\phi_h$ and $\phi_s$ are read off from the solutions at $r_h=1$, so Eq.~\eqref{eq:EO_numerical_working} can be evaluated directly.

\subsection{Low-frequency convergence of the Kubo extraction}
\label{app:omega_convergence}

\begin{figure}[H]
    \centering
    \makebox[\textwidth][c]{%
        \includegraphics[width=0.50\textwidth,trim=10pt 7pt 18pt 7pt,clip]{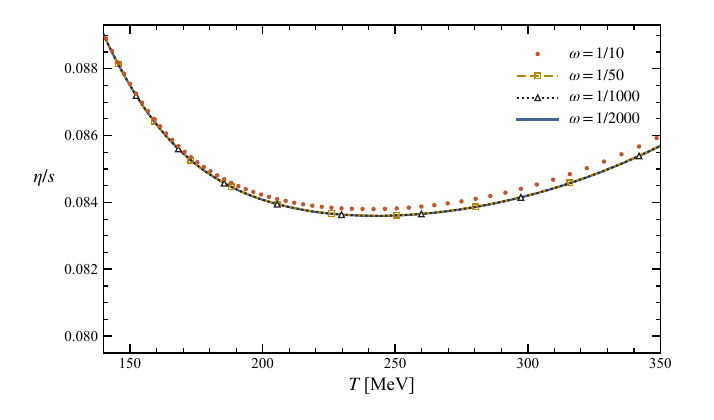}%
        \hspace{-0.005\textwidth}%
        \includegraphics[width=0.50\textwidth,trim=10pt 7pt 18pt 7pt,clip]{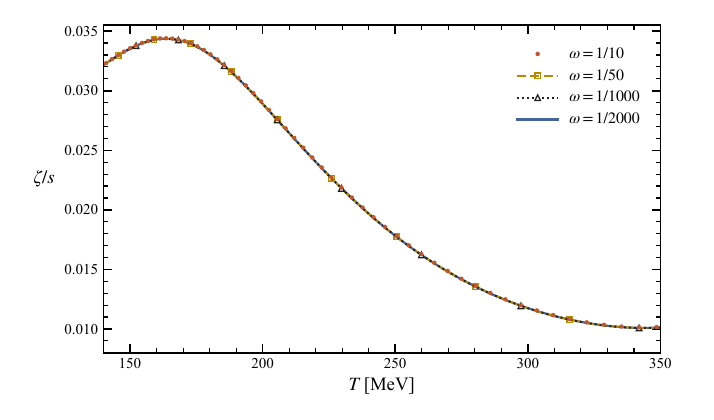}%
    }
   \caption{Low-frequency convergence of the boundary Kubo extraction for the scalar-dependent benchmark $H(\phi)=1-\phi^6/(1+2\phi^6)$ with $\alpha=-0.008$. The left and right panels show $\eta/s$ and $\zeta/s$, respectively, for four values of $\omega$ in units with $r_h=1$. The results at $\omega=1/1000$ and $\omega=1/2000$ coincide to graphical accuracy over the plotted temperature range.}

    \label{fig:omega_convergence}
\end{figure}

We evaluate the transport coefficients in Section~\ref{sec:horizon_validation} at a small finite frequency, $\omega=1/2000$, in units with $r_h=1$. To assess residual frequency dependence, we repeat the boundary Kubo calculation at several $\omega$ values. We perform this test for the scalar-dependent benchmark $H(\phi)=1-\phi^6/(1+2\phi^6)$. As shown in Figure~\ref{fig:omega_convergence}, finite-frequency corrections are visible at larger $\omega$. However, the results for $\omega=1/1000$ and $1/2000$ overlap at graphical resolution across all temperatures. This confirms the numerical stability of our extraction at low frequencies.

\bibliographystyle{JHEP}
\bibliography{references}

\end{document}